\documentclass[letterpaper,twocolumn,10pt]{article}
\usepackage[letterpaper,margin=1in,columnsep=0.33in]{geometry}
\usepackage{times}
\usepackage[T1]{fontenc}
\usepackage[utf8]{inputenc}
\usepackage{booktabs}
\usepackage{float}
\usepackage{multirow}
\usepackage{graphicx}
\usepackage{xcolor}
\definecolor{traceblue}{HTML}{245A81}
\definecolor{tracered}{HTML}{B53B38}
\usepackage{amsmath}
\usepackage{amssymb}
\usepackage{caption}
\usepackage{comment}
\usepackage{microtype}
\usepackage{tikz}
\usetikzlibrary{arrows.meta,positioning,fit,backgrounds,calc}
\usepackage[hidelinks]{hyperref}
\usepackage{xurl}
\usepackage{enumitem}
\setlist{nosep,leftmargin=1.2em}
\newcommand{\sys}{\textsc{BACGuard}}
\newcommand{\cve}[1]{\mbox{CVE-#1}}

\title{\vspace{-2em}\textbf{\sys{}: Fact-Grounded Runtime Mitigation of
Broken Access Control}}
\author{
    {\rm \textit{Moses Ndebugre}} \\
    North Carolina Agricultural and Technical State University \\
    Greensboro, North Carolina, USA
}
\date{}
  
\begin{document}
\maketitle

\begin{abstract}
Broken Access Control (BAC) has been the foremost category in the OWASP Top Ten since 2021, primarily due to the challenges in enforcing authorization at the application layer. Traditional mechanisms depend on logic embedded in the application, referencing relational database information such as permissions, roles, or ownership. However, missing or incorrect checks can go undetected by external systems, leaving applications vulnerable.

This paper presents \sys{}, a reverse proxy designed to enforce authorization for unmodified applications. \sys{} reconstructs authorization relations by leveraging authenticated user data from session stores and object ownership from relational schemas. It evaluates whether users are permitted to access resources and blocks unauthorized requests in real time. Unlike existing approaches, \sys{} bases its authorization on database joins, making the system resilient to forgery. For most requests, decisions are made deterministically. In cases where access control violations are only apparent in response data, such as cross-user data leakage, a language model is used as a fallback.

Evaluations on six applications, including those with known vulnerabilities, demonstrate that \sys{}'s hybrid deterministic and language model-based approach blocks all tested attacks with a false-positive rate below 0.7\%. Competing solutions like Web Application Firewalls are less effective and more disruptive. The results highlight \sys{}’s primary contribution: the external reconstruction of authorization facts, enabling robust enforcement even when the application itself is flawed.
\end{abstract}

\section{Introduction}
Broken Access Control (BAC) represents the most common type of vulnerability in web applications. In 2021, BAC became the leading category in the OWASP Top 10 due to its frequent occurrence across tested applications~\cite{owasptop10}. Variants at the object and function level, specifically Broken Object Level Authorization (BOLA) and Broken Function Level Authorization (BFLA), also top the OWASP API Security rankings~\cite{owaspapi}. A BAC vulnerability allows users to exceed their authorized permissions, such as accessing or altering other users' information or executing administrative operations. Because authorization mechanisms are distributed throughout application logic rather than enforced at a single point, a single missed check can compromise an entire set of objects.

Defending against BAC externally is particularly difficult because the vulnerability is semantic and specific to individual objects, rather than syntactic. Malicious and legitimate requests generally appear identical in structure, as both are well-formed and authenticated. The primary differentiator is object ownership, which is stored in the database. Web Application Firewalls (WAFs) that rely on request syntax analysis cannot discern this ownership distinction. For instance, the WAF baseline blocks only 40\% of attacks but also affects 25\% of legitimate traffic. Existing defenses often involve analyzing, instrumenting, or repairing application source code, limiting their applicability to certain deployed or third-party systems ~\cite{rolecast,fixmeup,nemesis,sun2011static}. This motivates investigating whether authorization facts can be reconstructed from existing application artifacts and enforced through an external proxy.

This work explores whether an external defense can be effective even without direct access to an application's internal concepts of identity and ownership. The central insight is that these concepts are not hidden, but are instead reflected in artifacts made available by the application. Identity is communicated via the authentication token or session store, while ownership is represented through foreign keys in the relational schema. \sys{} operates as a reverse proxy in front of an unmodified application, reading these artifacts in a read-only fashion and reconstructing the authorization checks that should have been enforced. The verified caller, true object owner, and the caller's administrative status are derived directly from the application's data, not from request parsing, which ensures both accuracy and protection against tampering. Ownership cannot be falsified if it is established through a database join.

This study addresses four research questions:
\begin{itemize}
\item \textbf{RQ1 (Detection).} To what extent can the system accurately detect BAC exploitation across PortSwigger BAC classes~\cite{portswigger-bac}, achieving high recall while minimizing false positives?
\item \textbf{RQ2 (Mitigation).} Does the system prevent attacks effectively, comparing attack success rates with and without the guard, while maintaining legitimate functionality?
\item \textbf{RQ3 (Mechanism).} What factors contribute to \sys{}'s effectiveness, such as the external reconstruction of authorization information or the agent topology, and how does its performance compare to that of a conventional WAF?
\item \textbf{RQ4 (Cost).} What are the added latency and API costs, and to what extent can a deterministic fast path mitigate these overheads?
\end{itemize}

The primary contributions of this work are outlined below.
\begin{enumerate}
    \item Introduction of runtime, fact-based BAC enforcement for unmodified applications, where each request is evaluated through a single relational check, permitted(principal, action, resource), over an authorization model reconstructed in a read-only manner from the application's data.
    \item Comprehensive evaluation across six applications and nine BAC classes, encompassing three real-world CVEs in production systems, using per-decision Wilson confidence intervals, and providing statistically distinct WAF and single-agent baselines.
    \item A two-phase adaptive attacker study that identifies and addresses two bypass classes in the first phase, and successfully resists leak-agent injection and automated fuzzing attacks in the second phase, demonstrating the architecture's resilience to the evaluated prompt injection attacks.
\end{enumerate}

\subsection{A motivating example}\label{sec:motiv}
For example, the endpoint \texttt{GET /books/v1/\{title\}} returns a book's contents. The proper authorization check requires confirming that the requester is the book's owner (\texttt{books.user\_id = session.user}). If a developer neglects this check, any authenticated user can access any book by specifying its title, resulting in an Insecure Direct Object Reference (IDOR), which exemplifies the object-level form of BAC (BOLA). The attacker simply issues a legitimate-looking request, such as \texttt{GET/books/v1/Learn\_Python}, using a valid token. The request's syntax, including the method, path, headers, and authentication, is indistinguishable from a legitimate request. The difference lies in the database, where the \texttt{user\_id} for the requested book is Alice's, not the attacker's, a detail not present in the request and undetectable by syntactic filters.

This example demonstrates why external defenses that rely solely on request matching are generally ineffective, but it also points towards a possible solution. Since the missing authorization check is fundamentally a database relationship, a guard that can access and interpret this relationship can reconstruct the appropriate authorization control.

\begin{figure}[t]\centering
\setlength{\fboxsep}{6pt}%
{\color{traceblue!65}
\fbox{%
\color{black}
\begin{minipage}{0.93\columnwidth}
\footnotesize\ttfamily\raggedright
\textcolor{traceblue}{\textbf{REQUEST}}\
GET /books/v1/Learn\_Python\ (valid JWT)\\[2pt]
\textcolor{traceblue}{(1) IDENTITY} (det.):
sig OK $\to$ caller="attacker"(7), admin=F\\
\textcolor{traceblue}{(2) ROUTE} (derived):
/books/v1/\{title\} $\to$ book:Learn\_Python,\\
\hphantom{(2) }sensitivity=owner\\
\textcolor{traceblue}{(3) OWNERSHIP} (det./DB):
books.user\_id$\to$users.id\\
\hphantom{(3) }$\Rightarrow$ owner="alice"\\
\hphantom{(3) }FACTS=\{caller:attacker, admin:F,\\
\hphantom{(3) FACTS=\{}owner:alice, caller\_owns:F\}\\
\textcolor{traceblue}{(4) DETECTOR} (LLM):
owner route, caller$\neq$owner, !admin\\
\hphantom{(4) }$\to$ bac\_attempt=T
("reads another's book")\\
\textcolor{traceblue}{(5) ARBITER} (LLM):
facts override policy $\to$ \{block, deny\}\\
\textcolor{tracered}{\textbf{RESPONSE}\ 403}\
X-BACGuard: \textcolor{tracered}{block}
(class=IDOR/BOLA)\\[3pt]
\normalfont\footnotesize
\emph{\textcolor{traceblue}{hybrid}}:
the ownership relation settles this with no LLM call.
\emph{\textcolor{traceblue}{injection}}:
a request body ``I own this''
cannot flip \texttt{caller\_owns},
as it is a DB join, not request text.
\end{minipage}}}
\caption{A worked single-request trace of the motivating IDOR.}
\label{fig:trace}
\end{figure}

\begin{figure}[h]
    \centering
    \includegraphics[width=\linewidth]{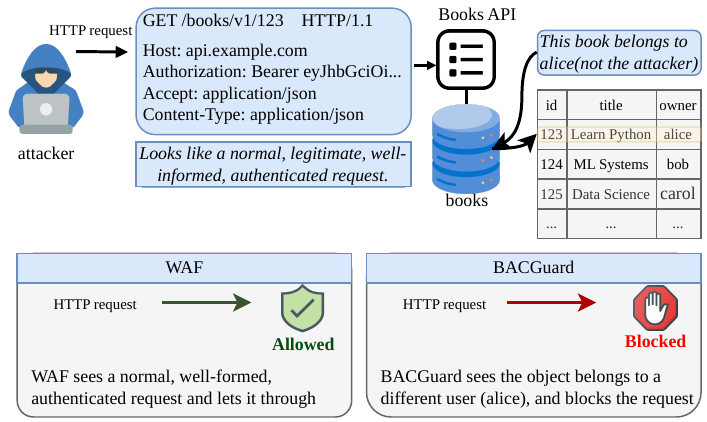}
    \caption{An IDOR that no request filter can catch.}
    \label{fig:motivatingexample}
\end{figure}
\section{Threat Model}

\begin{figure*}[t]
    \centering
    \includegraphics[width=\linewidth]{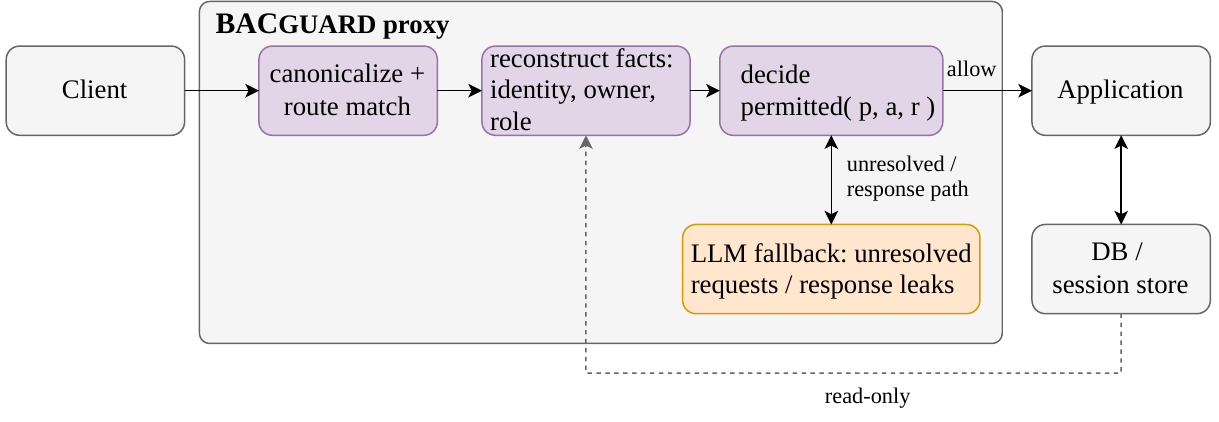}
    \caption{\sys{} architecture. Identity and ownership facts are obtained through read-only access to the application's data stores. A deterministic fast path evaluates the authorization predicate permitted(principal, action, resource). Requests unresolved by this path and response-path leakage checks may be delegated to the LLM. A block returns HTTP 403 to the client.}
    \label{fig:arch}
\end{figure*}

\hspace{1em} \textbf{Application:} The application is presumed honest but flawed, with endpoints lacking complete access-control checks or depending on client-supplied information, such as roles, visibility flags, or identifiers. Although the database accurately maintains ownership, membership, and roles, the application enforces these controls imperfectly. This situation commonly leads to broken access control vulnerabilities, where correct data exists but enforcement is inadequate. 

\textbf{Attacker:} The attacker may be authenticated or anonymous and seeks to access unauthorized resources by manipulating requests, forging tokens, or inserting adversarial data. Techniques include altering identifiers, spoofing values, and evading detection through request mutations. Network injection and remote code execution fall outside the scope of this model. 

\textbf{Guard:} \sys{} acts as a proxy on the request path, reading from the application's datastore and session store without altering the application. It relies on the same data as the application's ORM. If it cannot make a clear decision, the request is escalated to LLM agents, with refusals flagged as suspicious. The reliability of \sys{}'s identity verification depends on the integrity of trust anchors, such as token-verification keys. If a token secret is compromised, the proxy inherits this weakness. In cases where database access is unavailable, \sys{} can operate in a limited specification-only mode, as discussed in Section \ref{sec:deploy}.

\section{Design}\label{sec:design}
Requests are processed sequentially by the client, \sys{} proxy, application, and database, as depicted in Figure~\ref{fig:arch}. The orchestrator operates through the following stages: Identity verification using authentication tokens, access policy determination via static or dynamically derived specifications, ownership determination from the database, exploitation attempt detection using an LLM, and final decision-making by the arbiter. Additionally, a leak agent checks if responses expose another user's data. Figure~\ref{fig:trace} shows an example request trace.

\vspace{1em}
\noindent \textbf{The agents:} Each agent has a distinct function and uses a fixed prompt to generate structured outputs. The policy agent determines the sensitivity of routes, caching results for efficiency. The detector identifies BAC attempts by evaluating verified identities, access policies, and ground-truth facts. The arbiter integrates these signals to allow or block requests and suggests mitigation strategies. The leak agent identifies whether responses inadvertently reveal data of other users. Dividing responsibilities among specialized agents enhances both precision and resistance to injection compared to a single agent model, though the accuracy fundamentally relies on the quality of reconstructed facts.

\noindent \textbf{Fact grounding:} Identity, ownership, and administrative status are deterministically derived from application data for LLM agent evaluation. The arbiter prioritizes factual data over permissive policies because such policies can cause BAC vulnerabilities. Grounding authorization decisions in verified application data reduces reliance on attacker-controlled request content. This design limits opportunities for prompt injection to override established facts, although LLM-mediated decisions remain susceptible to reasoning errors and adversarial inputs.

\noindent \textbf{The authorization decision:} Authorization is determined by evaluating whether the principal is allowed to perform a given action on a resource, using the \(\operatorname{permitted}(p,a,r)\) predicate. This approach supports various authorization models, including roles and ownership, by reconstructing facts from identity and data stores. While general, the model may require additional policy information for complex scenarios. BACGuard handles cases with insufficient facts through predefined resolution strategies.

\noindent \textbf{Request resolution:} Prior to evaluation, \sys{} clarifies the principal, action, and resource denoted by each request. Requests with ambiguous or duplicate authorization-relevant parameters are rejected to ensure precise resource identification. This process is distinct from authorization logic.

\noindent \textbf{Request canonicalization:} \sys{} standardizes requests by percent-decoding, collapsing duplicate or trailing slashes, and normalizing numeric identifiers. This ensures consistency with backend processing and prevents bypasses due to path mutations.

\begin{figure}[h]
    \centering
    \includegraphics[width=\linewidth]{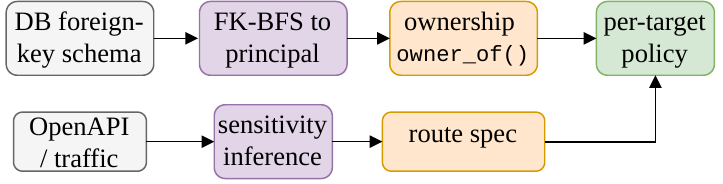}
    \caption{Auto-derived policy: ownership from the FK schema and the route spec from the OpenAPI document or observed traffic, with no policy being hand-written.}
    \label{fig:auto}
\end{figure}

\noindent \textbf{Auto-deriving the per-target policy:} Rather than manually defining per-target policies, \sys{} automatically generates the ownership model and route specification from the application's database schema and documentation, as illustrated in Figure~\ref{fig:auto}. The generator identifies ownership relations through foreign-key analysis and route sensitivity by referencing OpenAPI documents or observed traffic. Only the identity trust anchor must be provided manually, as it represents a credential.

\section{Implementation}
The proxy is implemented as a mitmproxy addon \cite{mitmproxy}. Agents communicate with an LLM API using structured JSON responses. Guard decisions are recorded as JSONL logs, which are used for later analysis.

Data stores are snapshotted in read-only mode, with snapshots cached briefly and refreshed if new objects are missing. The evaluation uses a variety of real-world identity resolution and authorization models, as summarized in Table~\ref{tab:targets}, demonstrating the guard's flexibility.

A per-decision cache, based on route, identity, ownership, and privilege, reduces redundant processing. The cache is reset for each experiment, enabling accurate LLM evaluation. The guard fails safely by escalating unresolved routes, treating agent refusals or parse errors as suspicious, and canonicalizing requests before matching.

\subsection{Experimental Setup}
The experiments use six applications, including three intentionally vulnerable ones to represent diverse authorization and identity schemes, and three production systems with known CVEs to ensure external validity, as shown in Table~\ref{tab:targets}.

\begin{table}[t]\small\centering
\caption{Overview of experimental targets: six technology stacks, six distinct authorization models, including three applications with real-world CVEs.}
\label{tab:targets}
\footnotesize
\begin{tabular}{@{}lll@{}}
\toprule
Target & Stack & Authz.\ model \\
\midrule
VAmPI \cite{vampi} & Flask/SQLite & JWT HS256 \\
Juice Shop \cite{juiceshop} & Node/SQLite & JWT RS256 \\
DVWA \cite{dvwa} & PHP/MySQL & session cookie \\
Grafana 8.1.5\,(CVE) \cite{cve-grafana} & Go/SQLite & capability key \\
Kanboard 1.2.29\,(CVE) \cite{cve-kanboard} & PHP/SQLite & object-ref.\ check \\
Gitea 1.21.1\,(CVE) & Go/SQLite & acct.\ visibility \\
\bottomrule
\end{tabular}
\end{table}

The evaluation examines nine types of broken access control, including IDOR and BOLA, broken function-level authorization, privilege escalation by mass-assignment, unprotected features, data leakage through response paths, manipulation of roles or IDs in requests, HTTP parameter pollution, and object-reference consistency issues.
\section{Results and Discussion}
\label{sec:eval}

\subsection{Answering RQ1, RQ2 and RQ3}
\begin{figure*}[t]\centering
\includegraphics[width=\linewidth]{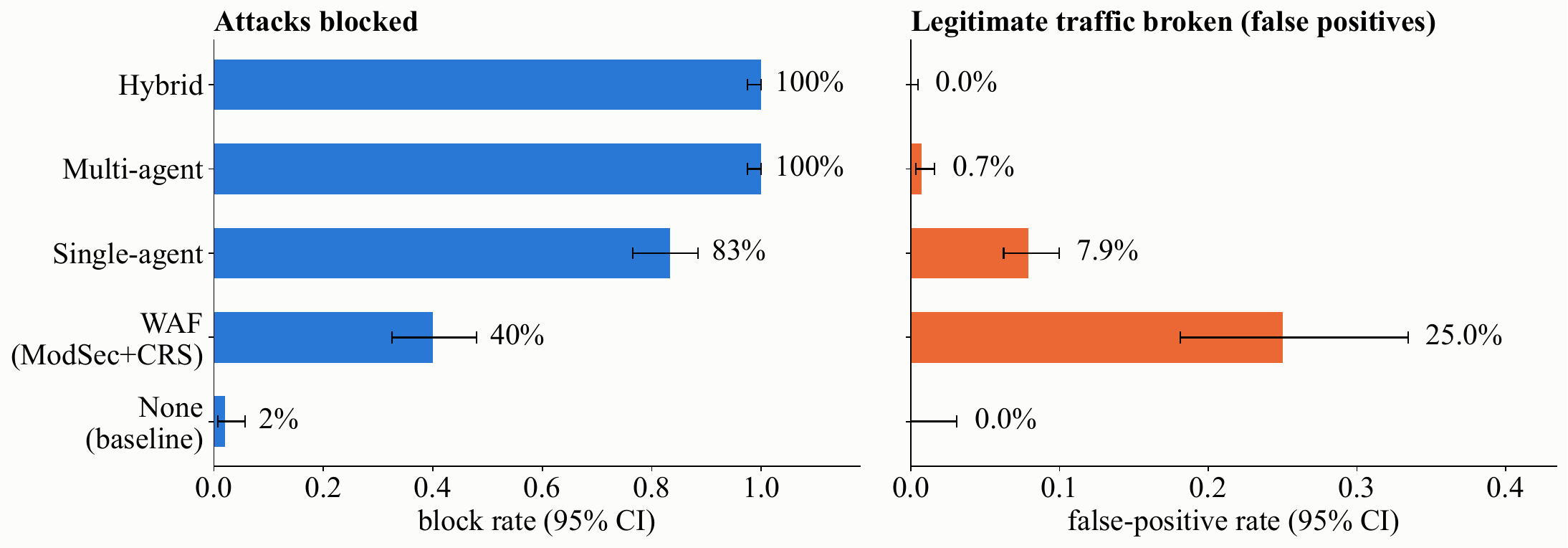}
\caption{Attacks blocked and false positives per defense (VAmPI, N=30, 95\% CI).}
\label{fig:rq3}
\end{figure*}

Table~\ref{tab:rq3} and Figure~\ref{fig:rq3} compare defense strategies on VAmPI. In the absence of protection, 98\% of attacks succeed. Multi-agent and hybrid methods block all attacks and keep false positives at or below 0.7\%. The web application firewall (WAF) stops only 40\% of broken access control attacks and causes a 25\% false positive rate, due to its reliance on generic rather than access-control-specific rules. The single-agent approach blocks 83\% of attacks but fails to prevent mass-assignment escalation and produces a 7.9\% false positive rate.

 Wilson intervals are computed over decisions, with $N{=}30$ repeats reflecting repeated measures of a fixed scenario set \cite{wilson1927}. Each VAmPI run includes five attack scenarios and 27 benign requests; per-target runs contain fewer scenarios. These intervals indicate consistency across repeated model sampling but do not measure generalization. Generalization is demonstrated by evaluating six authorization models, a blinded seventh application, and a 180-probe adversarial set.
 
\begin{table}[h]\small\centering
\caption{RQ3 on VAmPI (N=30, Wilson 95\% CI). Attacks $n{=}150$; benign
$n{=}120$ (baseline/WAF) or $810$ (agent rows).}
\label{tab:rq3}
\resizebox{\columnwidth}{!}{%
\begin{tabular}{@{}lcc@{}}
\toprule
Defense & Attacks blocked & Benign broken (FP) \\
\midrule
None & 0.02 [.01,.06] & 0.00 [.00,.03] \\
WAF (ModSec+CRS) & 0.40 [.33,.48] & 0.25 [.18,.33] \\
Single-agent & 0.83 [.77,.88] & 0.079 [.06,.10] \\
Multi-agent & 1.00 [.98,1] & 0.007 [.00,.02] \\
Hybrid & 1.00 [.98,1] & 0.00 [.00,.005] \\
\bottomrule
\end{tabular}}
\end{table}

\subsection{Answering RQ4}
\begin{figure}[t]\centering
\includegraphics[width=\linewidth]{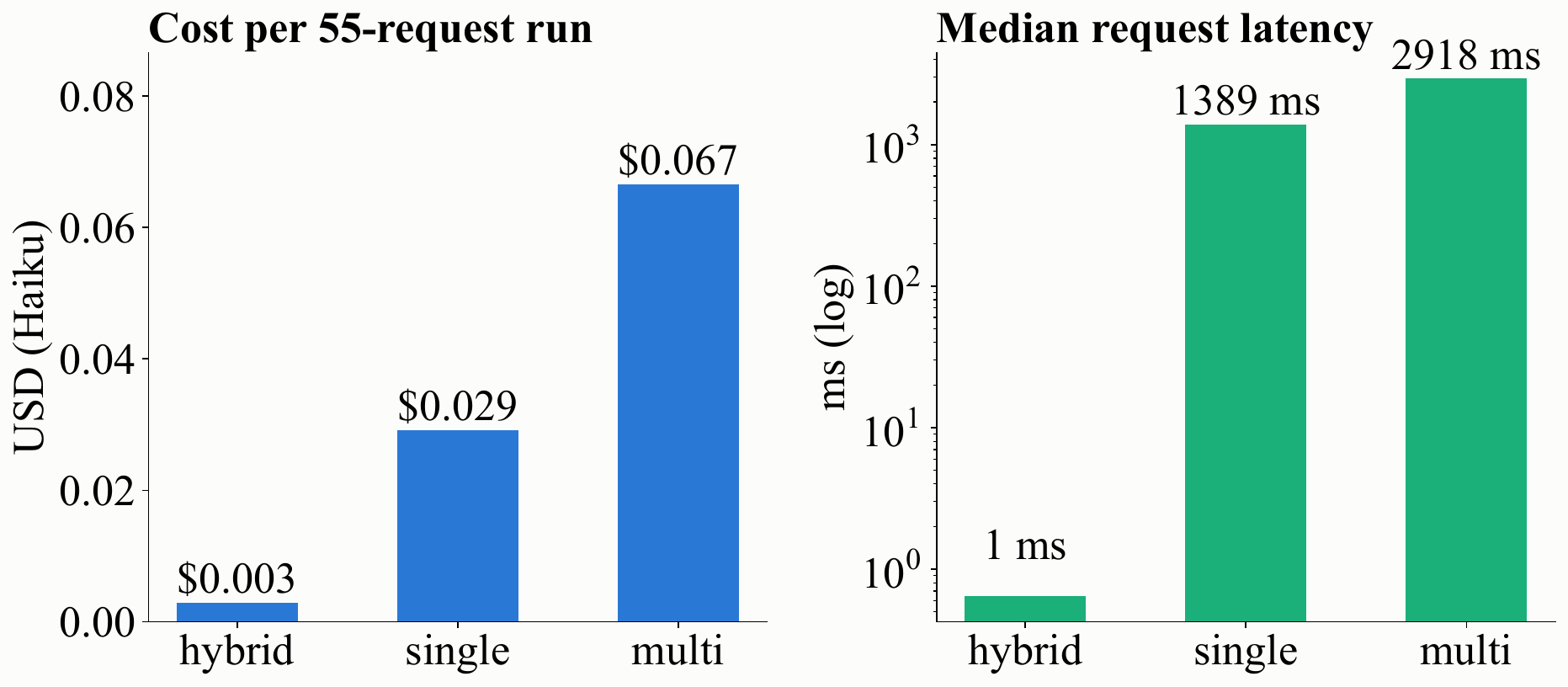}
\caption{Cost and median request latency by mode (VAmPI, N=30).}
\label{fig:rq4}
\end{figure}

In a 55-request evaluation (Figure.~\ref{fig:rq4}), the hybrid approach required about 3.9 LLM calls per run (\$0.003 and median latency of roughly 1 ms), compared to 32.7 for the single-agent (\$0.029 and 1.39 s) and 65 calls per run for the multi-agent (\$0.067 and 2.9 s), using Haiku pricing. The hybrid strategy matches multi-agent detection accuracy at approximately one-tenth of the cost and latency by using a deterministic path for clear cases and reserving LLM invocation for ambiguous ones. As traffic increases, the hybrid's per-request cost nears that of the deterministic method because the LLM is needed for no more than 12\% of requests, whereas other modes rely on the LLM for every request. Latency figures reflect guard-decision time; total overhead also includes database snapshot retrieval. The monetary figures here and in Table~\ref{tab:sweep} are from distinct workloads and are not directly comparable. The main finding is that the hybrid approach consistently ranks as the most cost-effective, achieving an approximate tenfold reduction in cost compared to pure-LLM approaches by limiting model use to a subset of requests.

\subsection{Generalization across six targets}
With per-target adapters, the reasoning engine blocks all attacks across six targets with a lower-bound confidence interval of at least 0.94 and a false positive rate at or below 0.7\% for multi-agent and hybrid modes (see Table~\ref{tab:cross}). The single-agent approach exhibits a class-dependent recall gap, missing only VAmPI's mass-assignment escalation. Hybrid and multi-agent modes outperform the single-agent baseline in recall and false positive rates. The improvements are primarily due to deterministic fact reconstruction and selective LLM invocation inherent to the hybrid design, rather than agent decomposition alone.

A 40-request browser session and a role-tagged discovery window replayed through the guard yielded zero false positives. Large-scale testing with thousands of requests confirmed these outcomes. Table~\ref{tab:cross} adds crAPI as a held-out target, where all five BOLA attacks were blocked without false positives, consistent with results from the other targets.

\begin{table}[t]\small\centering
\caption{Per-target block and false positive rates with 95\% confidence intervals. The held-out crAPI row ($\dagger$) is reported as raw counts from a single run. Multi/single modes were not evaluated on crAPI.}
\label{tab:cross}
\begin{tabular}{@{}lccc@{}}
\toprule
Target & multi & hybrid & single \\
\midrule
VAmPI & 1.00 / .007 & 1.00 / .00 & .83 / .079 \\
Juice Shop & 1.00 / .00 & 1.00 / .00 & 1.00 / .065 \\
DVWA & 1.00 / .00 & 1.00 / .00 & 1.00 / .00 \\
Grafana\,(CVE) & 1.00 / .00 & 1.00 / .00 & 1.00 / .00 \\
Kanboard\,(CVE) & 1.00 / .00 & 1.00 / .00 & 1.00 / .00 \\
Gitea\,(CVE) & 1.00 / .00 & 1.00 / .00 & 1.00 / .00 \\
\midrule
crAPI\,(held-out)$^\dagger$ & --- & 5/5 / 0/1 & --- \\
\bottomrule
\end{tabular}
\end{table}

\subsection{Coverage across BAC classes}
Table~\ref{tab:classes} links nine BAC classes to their corresponding targets, the guard mechanism used, and the decision method. Eight classes are resolved deterministically after route specification is set. Only the response-path data-leakage class requires BACGuard's LLM-based leak inspector, though alternative response-inspection methods are feasible.

\begin{table}[h]\small\centering
\caption{BAC classes, where exercised, and how \sys{} decides.}
\label{tab:classes}
\resizebox{\columnwidth}{!}{%
\begin{tabular}{@{}llc@{}}
\toprule
BAC class & Guard mechanism & Decide \\
\midrule
IDOR/BOLA (id/title/basket) & ownership (FK) & det. \\
Broken function-level auth & role & det. \\
Priv.\ esc.\ (mass-assign) & privilege-field & det. \\
Unprotected functionality & role/visibility & det. \\
User-ID/role via request & identity vs.\ ground truth & det. \\
HTTP parameter pollution & sanitizer & det. \\
Object-ref.\ consistency & FK context match & det. \\
Account-visibility leak & visibility (DB flag) & det. \\
Response-path data leakage & leak agent (LLM) & \textbf{LLM} \\
\bottomrule
\end{tabular}}
\end{table}

\subsection{Real-world CVEs}
Grafana \cve{2021-39226}~\cite{cve-grafana} involves information leakage due to enumerable references. \sys{} mitigates this by treating snapshots as bearer-capability objects and allowing access only with valid keys from the database. For Kanboard \cve{2023-33956}~\cite{cve-kanboard}, \sys{} checks the file-to-project association and blocks mismatched requests. In Gitea \cve{2025-68945}, \sys{} ensures only authorized users can access private resources. All three CVEs were reproduced and successfully blocked in all modes without false positives.

\subsection{Model vs.\ architecture}
A per-model sweep with $N{=}30$ (Table~\ref{tab:sweep}) indicates that all models block at least 149 out of 150 attacks in every mode, making false positives and cost the main factors of distinction. While stronger models improve single-agent performance, the hybrid design achieves similar or superior results with smaller models at much lower cost. This improvement stems from deterministic fact reconstruction and selective LLM invocation, rather than the number of agents involved.

Notably, Opus is the only model with false positives in hybrid mode (3 out of 600), due to an over-eager response-path leak agent. In contrast, Haiku and Sonnet hybrids have zero false positives. The hybrid fast path maintains a latency of approximately 1 ms by invoking the model only for response-path decisions.

\begin{table}[h]\centering\footnotesize
\caption{Per-model sweep on VAmPI, $N{=}30$ (150 attack / 600 benign decisions per
cell). Detection saturates; FP and cost discriminate. ``\$/run'' is the mean API
cost of one full battery run.}
\label{tab:sweep}
\resizebox{\columnwidth}{!}{%
\begin{tabular}{@{}llcccc@{}}
\toprule
model & mode & block & FP & \$/run & lat. \\
\midrule
\multirow{3}{*}{Haiku}
 & single & 99.3\% & 10.7\% (64/600) & 0.025 & 1464\,ms \\
 & multi  & 100\%  & 1.7\% (10/600)  & 0.056 & 3231\,ms \\
 & hybrid & 100\%  & \textbf{0\% (0/600)} & 0.002 & \textbf{0.7\,ms} \\
\midrule
\multirow{3}{*}{Sonnet}
 & single & 100\% & 0.5\% (3/600) & 0.070 & 2233\,ms \\
 & multi  & 100\% & \textbf{0\% (0/600)} & 0.147 & 4510\,ms \\
 & hybrid & 100\% & \textbf{0\% (0/600)} & 0.006 & 0.7\,ms \\
\midrule
\multirow{3}{*}{Opus}
 & single & 100\% & 0.5\% (3/600) & 0.131 & 2804\,ms \\
 & multi  & 100\% & 0.5\% (3/600) & 0.294 & 6422\,ms \\
 & hybrid & 100\% & 0.5\% (3/600) & 0.010 & 0.7\,ms \\
\bottomrule
\end{tabular}}
\end{table}

%%%%%%%%%%%%%%%%%%%%%%%%%%%%%%%%%%%%%%%%%%%%
The VAmPI evaluation was conducted with agents running on free, downloadable open-weight models, specifically Qwen2.5-Coder-7B-Instruct and Qwen2.5-14B-Instruct. The multi-agent setup showed improved false-positive reduction compared to the single-agent baseline. However, this improvement cannot be solely attributed to agent decomposition, as other factors such as fact grounding and decision processes may contribute. The hybrid configuration was the most efficient, resolving most requests deterministically and invoking the LLM only when needed.

\begin{table}[h]\centering\footnotesize
\caption{\sys{} on free, local open-weight models (VAmPI, \$0 API).}
\label{tab:local}
\resizebox{\columnwidth}{!}{%
\begin{tabular}{@{}llccc@{}}
\toprule
model & mode & block & FP & median lat. \\
\midrule
\multirow{3}{*}{Qwen2.5-Coder-7B}
 & single & 15/15 (100\%) & 30/60 (50.0\%) & 2868\,ms \\
 & multi  & 10/10 (100\%) & 11/40 (27.5\%) & 7733\,ms \\
 & hybrid & 15/15 (100\%) & \textbf{0/60 (0\%)} & \textbf{0.7\,ms} \\
\midrule
\multirow{3}{*}{Qwen2.5-14B}
 & single & 12/15 (80\%)  & 5/60 (8.3\%)  & 4792\,ms \\
 & multi  & 10/10 (100\%) & \textbf{0/40 (0\%)}  & 14267\,ms \\
 & hybrid & 15/15 (100\%) & \textbf{0/60 (0\%)} & \textbf{0.7\,ms} \\
\bottomrule
\end{tabular}}
\end{table}

\begin{figure}[h]\centering
\includegraphics[width=\linewidth]{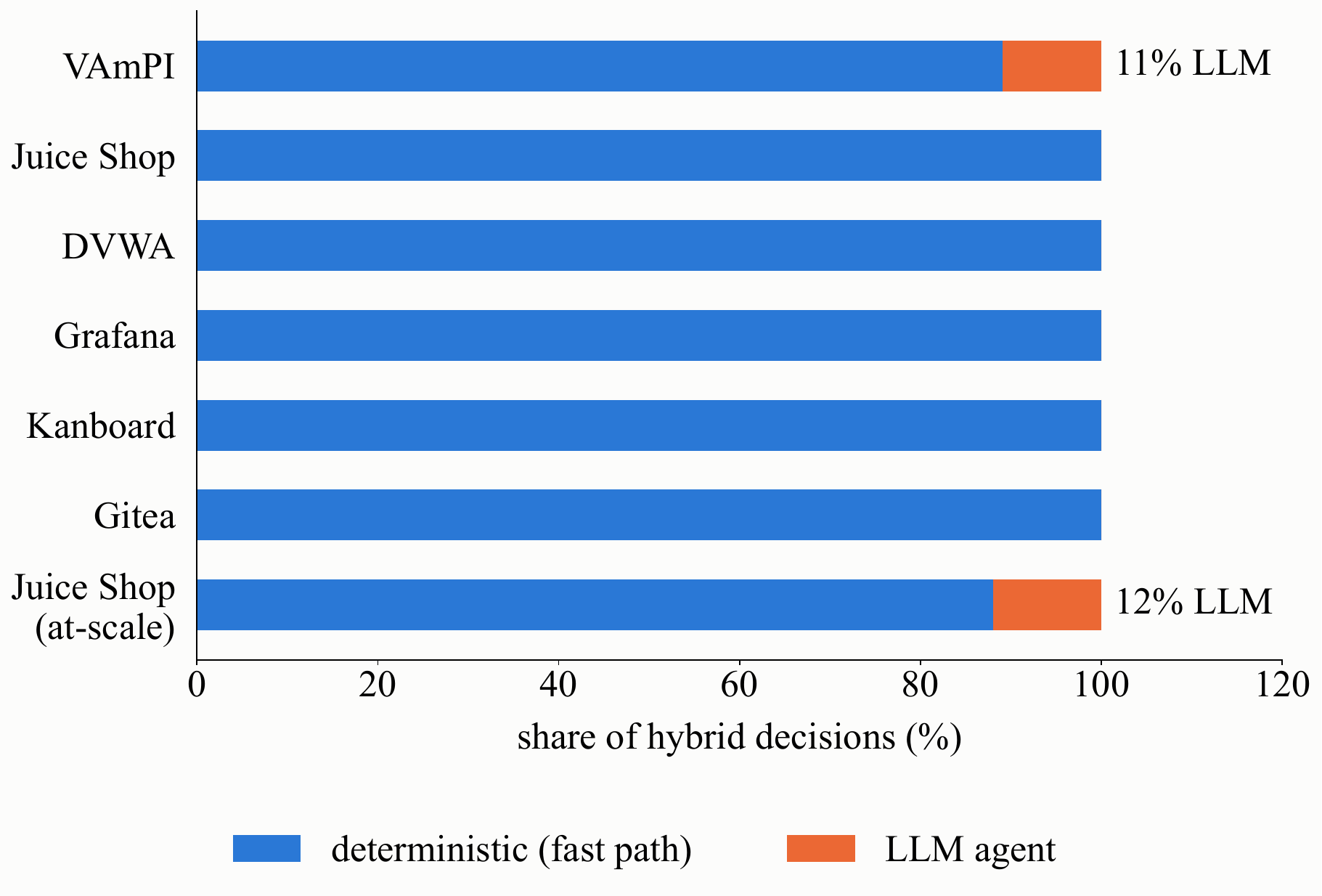}
\caption{Share of hybrid decisions resolved deterministically vs. delegated to an LLM agent. The deterministic path dominates; the LLM fires only on the response-path class (VAmPI) and routes outside the spec (Juice Shop at-scale).}
\label{fig:deleg}
\end{figure}

\subsection{Benign workloads: synthetic, captured, and browser}
To evaluate deployability, three categories of benign input were tested. A hand-crafted set of legitimate actions, a replayed Juice Shop session, and browser automation using Playwright/Chromium all resulted in zero false positives for both multi-agent and hybrid modes. The browser-based test identified endpoints not included in synthetic datasets, highlighting the importance of real client traffic in revealing edge cases where overly strict guards might block valid activity.

To test scalability, a traffic generator simulated various user roles and sessions using the hybrid deployment. Out of 1,732 benign requests across three targets, there were no blocks, yielding a 95 percent confidence upper bound of 0.22 percent for the false positive rate.

\subsection{What the LLM actually contributes}
\label{sec:llmvalue}
An important question is whether the LLM is necessary or if deterministic logic alone suffices for effective guarding. Analysis shows that in hybrid mode, 88 to 90 percent of decisions are made deterministically, with only a small fraction requiring the LLM, primarily for leak checks or unspecified routes. In the ablation, deterministic request-stage enforcement prevented the evaluated request-stage attacks but did not detect VAmPI's response-path leakage case. \sys{}'s LLM-based leak inspector detected this case, demonstrating the value of response inspection without establishing that an LLM is the only viable approach.

{\begin{table}[t]\small\centering
\caption{Block and false positive rates per agent, along with request-path LLM fallback frequency. Hybrid-single matches hybrid; only response-path leaks on VAmPI activate the model.}
\label{tab:isolation}
\resizebox{\columnwidth}{!}{%
\begin{tabular}{@{}lccc@{}}
\toprule
Target & deterministic & hybrid-single & hybrid \\
\midrule
VAmPI      & 4/5\ \ 0 FP & 5/5\ \ 0 FP & 5/5\ \ 0 FP \\
Juice Shop & 3/3\ \ 0 FP & 3/3\ \ 0 FP & 3/3\ \ 0 FP \\
DVWA       & 3/3\ \ 0 FP & 3/3\ \ 0 FP & 3/3\ \ 0 FP \\
Grafana    & 1/1\ \ 0 FP & 1/1\ \ 0 FP & 1/1\ \ 0 FP \\
Kanboard   & 1/1\ \ 0 FP & 1/1\ \ 0 FP & 1/1\ \ 0 FP \\
Gitea      & 1/1\ \ 0 FP & 1/1\ \ 0 FP & 1/1\ \ 0 FP \\
\midrule
req-path LLM fallbacks & --- & 0 & 0 \\
\bottomrule
\end{tabular}}
\end{table}

\subsection{Agent isolation: an equally-informed single agent}
\label{sec:isolation}
To address concerns that the benefits of the multi-agent design result from a less informed single agent, a hybrid-single mode was tested. This configuration merges the detector and arbiter into one agent but retains all other conditions of the hybrid setup. Testing showed that hybrid-single and hybrid modes performed identically in both attack blocking and false positive avoidance, indicating no inherent detection advantage from agent decomposition under these conditions.

This outcome occurs because, in these runs, the request-path LLM fallback is never triggered: the deterministic fast path resolves every request-stage case after fact reconstruction, so only VAmPI's response-path leak check involves the model. The held-out crAPI (\S\ref{sec:heldout}) further confirms this finding: its BOLA decision is made on the request path with no model call. Results show that multi-agent decomposition does not provide a detectable request-stage detection advantage over an equally informed single agent under the tested conditions. Improvements over the monolithic baseline (Table~\ref{tab:rq3}) stem from the hybrid design, which combines deterministic fact reconstruction, rule-based enforcement, and selective LLM invocation. As the request-path fallback was not triggered in this comparison, the specific contribution of detector–arbiter decomposition could not be isolated.

\subsection{An equally-informed deterministic baseline (OPA)}
\label{sec:opa}
When the request-stage decision uses only reconstructed facts, a standard policy engine such as Open Policy Agent reproduces the same blocking and false positive rates as \sys{}'s deterministic logic. This indicates that the core contribution at this stage is fact reconstruction, not a new decision mechanism. Both deterministic request-stage approaches miss VAmPI's response-path leakage case, which BACGuard's LLM-based inspector successfully detects. This result demonstrates the limitation of request-stage enforcement in the evaluated scenario, rather than proving that response-level detection requires an LLM.

%%%%%%%%%%%%%%%%%%%%%%%%%%%%%%%%%%%%

\subsection{Does the inferred policy match intended permissions?}
\label{sec:policyinfer} 
The OPA control demonstrates that decisions are simple when facts are correct, but it does not verify policy accuracy. An independent oracle evaluates each route using four types of principals: anonymous, authenticated non-owner, owner, and admin. Sensitivity is determined by whether the least-privileged principal receives a 2xx response, serving as ground truth for routes with enforced access control. In VAmPI and Juice Shop, auto-derived sensitivity matches enforcement for all controlled routes. Across 13 routes, \sys{} does not infer excessively strict policies. The other 7 routes are under-enforced, but \sys{} matches the documented policy and addresses the relevant gap. For instance, Juice Shop's \texttt{GET /api/Users} exposes 1,590 user records to regular users. The primary limitation is the small sample size, but the inferred policy aligns with available signals and avoids unnecessary restrictions.

\subsection{Ownership is not authorization: the abstain path}
\label{sec:abstain}
Ownership does not always equate to authorization, as real-world schemas often support access through sharing tables, delegation, or multiple foreign-key paths. Owner-only rules may exclude legitimate shared access and misinterpret relationships. To address this, the ownership resolver classifies principals as clear, conditional, or provides a reason, abstaining when authorization depends on ambiguous relationships. In synthetic schema tests with complex sharing and conflicting foreign keys, owner-only rules blocked 3 of 5 legitimate accesses, while the fact-based resolver blocked none. The resolver permits clear cases, denies unauthorized access, and abstains when permission semantics are unclear, deferring such decisions to the application or further review.

Synthetic cases demonstrate the impact of sharing relationships, which is also observed in a live application. In Kanboard, a task created by an admin is shared with one user and not with another, as determined by the application's join table. An owner-only rule incorrectly blocks the legitimate project member. In contrast, the fact-based resolver correctly permits both the owner and the shared member, while blocking only the non-member, resulting in no false blocks. Both types of relationships are derived from the application's tables.

crAPI exemplifies this issue in practice. Its mechanic-report endpoint is accessible to both the vehicle owner and the assigned mechanic, each identified via distinct foreign-key paths. A single-path resolver would fail to recognize one principal, but the revised approach authorizes both. Consequently, only unauthorized users are denied, and both legitimate roles are permitted, confirming the multi-path ambiguity observed in synthetic cases.

\subsection{Tail latency under concurrent load}
\label{sec:tail}
Median values do not represent the full impact of delegated requests. In testing, 300 concurrent requests, with 10 percent directed to the response-path or LLM branch, were processed through both the hybrid guard and directly to the target. Fast-path requests remained efficient, but those escalated to the LLM had a median response time of about 2.0 seconds, increasing the overall p95 and p99 to nearly 2 seconds, while the median request time was 27.7 milliseconds. Throughput decreased from 232 to 42 requests per second under the guard due to the LLM branch. These findings support the hybrid design, emphasizing that restricting model use to response-path cases is necessary for deployment. Systems with stringent tail latency requirements should limit or handle delegated requests asynchronously.

\subsection{Auto-derived policy}
\label{sec:autoderive}
Applying the same generator to three schemas, the system recovers VAmPI, Juice Shop, and Kanboard ownership relations. Each guard, reconfigured to use auto-derived ownership, replicates hand-written detection with all attacks blocked and no false positives. Where OpenAPI documentation exists, the derived specifications match the hand-written versions. For instance, \texttt{DELETE\,/users/v1/\{username\}} is correctly inferred as admin only. The automated Juice Shop guard, using traffic routes and foreign-key ownership without OpenAPI or hand-written policy, also blocks all attacks with zero false positives. Compared to hand-written specifications, it achieves complete route recall and sensitivity accuracy, with lower precision due to additional, valid routes absent from the minimal hand-written specification.

\subsection{Blind held-out evaluation (crAPI)}
\label{sec:heldout}
To evaluate cross-application generalization without target-specific policy configuration, BACGuard was tested on OWASP crAPI, which had not been used during development. The focus was on a previously unseen BOLA-vulnerable vehicle-location endpoint rather than on multiple vulnerability classes. crAPI is distinct from the primary targets due to its RS256 or JWKS authentication and PostgreSQL-based ownership relationships.

BACGuard automatically derived the vehicle ownership relationship from the live PostgreSQL schema. Five unauthorized requests targeting different users' vehicle UUIDs and one legitimate owner request were tested. The unguarded application disclosed all victims' locations, while BACGuard blocked the unauthorized requests and allowed the legitimate one. All decisions were made deterministically, without using an LLM.

These results show that the fact-reconstruction mechanism transfers to an unseen application, database backend, and authentication scheme, rather than generalizing across different BAC classes. Additional shared-access and multi-path cases evaluated on crAPI and Kanboard further support the approach for complex authorization relationships. A broader evaluation across multiple endpoints is left for future work.

\subsection{Adaptive-attacker analysis}
\label{sec:adaptive}
A 22-probe evaluation examined path normalization, ID normalization, HTTP parameter pollution, verb override, URL-rewrite headers, and JWT forgery, identifying two bypasses. Path normalization with duplicate slashes led to missed routes and unintended public access, which was resolved through request canonicalization. HTTP parameter pollution, where PHP and the matcher interpret parameters differently, enabled profile access, but this was fixed by blocking duplicated, authorization-relevant parameters. After these corrections, no evasions occurred in any probe. Identity forgery attempts were rejected except for tokens signed with a weak secret, an application-level issue. Prompt injection attempts were consistently blocked due to reliance on deterministic ownership facts. Fact grounding provides structural protection against injection. Round 1 used a single model and limited payloads, with two remaining probes evaluated below.

The two remaining vulnerabilities from round 1 were then assessed. For response-reflected injection targeting the Leak agent, adversarial text was included in the response body to induce false negatives. In all six payloads, the guard blocked every attempt, with no evasions. The Leak agent receives verified caller information separately, preventing ownership reassignment through the response body, in line with the fact-grounding approach. For automated path and parameter fuzzing, the canonical VAmPI IDOR was mutated across various dimensions, resulting in 30 variants. None of these variants bypassed the guard, as canonicalization and the sanitizer neutralized the tested mutations.

While rounds 1 and 2 focused on VAmPI, the analysis was extended to other authorization models. Each target's canonical IDOR or BOLA attack was replayed through the guard using a range of mutations, including duplicate and leading slashes, dot-segments, percent-encoding, case changes, trailing junk, semicolon path parameters, null bytes, Unicode confusables, and identity-injection headers. Each variant replicated the target's actual attack and verification method, with evasion defined as a successful, unblocked attempt confirmed by the oracle. Across six targets and 180 probes, no evasions were observed. Canonicalization blocked path-mutation attacks, the sanitizer excluded ambiguous parameters, and injected identity claims could not override deterministically derived ownership facts.

\section{Deployment Considerations}\label{sec:deploy}
\textbf{Guard Placement.} \sys{} serves as a proxy, acting as a reverse proxy for production deployments and a forward proxy in experiments. When deployed as a sidecar with one guard per application replica, each sharing a read replica of the database, \sys{} preserves local access to ground-truth data and avoids central bottlenecks.

\noindent \textbf{Handling Unreadable Databases.} The analysis assumes read-only access to the application's data store. If this is not possible, \sys{} maintains partial functionality by verifying identity through token validation. OpenAPI security scopes continue to define sensitivity levels. Object-reference and capability checks can be achieved via the application's API, though object-ownership checks require either a read replica or a minimal ownership endpoint. Improving this degraded mode is left for future work.

\noindent \textbf{Fail-Open Versus Fail-Closed.} In deterministic mode, \sys{} defaults to a fail-open behavior on unresolved routes to avoid disrupting unknown but harmless endpoints. In hybrid mode, these routes are escalated to the LLM, which handled all cases correctly in evaluation. Operators may also select a fail-closed policy for unspecified routes, accepting a temporary risk of false positives until route discovery is complete.

\section{Related Work}
\subsection{Detecting access-control bugs.}

Prior research focuses on identifying missing or incorrect authorization checks within applications, utilizing methods such as inferring proper check locations~\cite{rolecast}, repairing bugs~\cite{fixmeup}, detecting privilege escalation~\cite{mace}, and enforcing access control through dynamic information-flow tracking~\cite{nemesis}. Static and dynamic analyses have been used to detect per-role access issues~\cite{sun2011static}, logic vulnerabilities~\cite{felmetsger2010logic}, and session invariants~\cite{swaddler2007}, while tools like Baaz~\cite{baaz2010} and NAVEX~\cite{navex2018} address misconfigurations and validate flaws. These approaches interact directly with the application. In contrast, \sys{} is deployed externally as a runtime guard for unmodified applications, deriving correct authorization from the application's ground truth, such as databases or session stores, and actively mitigating issues by blocking unauthorized requests.

Recent research addresses BAC through detection and dynamic analysis. ABSENTIA~\cite{duarte2026absentiadetectingbrokenaccess} uses a large-language-model agent to audit application routes and infer necessary authorization properties, highlighting missing checks. It has demonstrated that pattern-based analyzers like CodeQL and Semgrep fail to detect these issues, emphasizing that authorization functions as a relation rather than a data flow traceable by signatures.

Dynamic approaches such as AuthProbe~\cite{barach2026authprobespecificationdrivenmultiidentitydetection}, BACScan~\cite{10.1145/3719027.3744825}, and BACFuzz~\cite{dharmaadi2025bacfuzzexposingsilencebroken} test applications under multiple identities to identify unauthorized access, while BOLARAY~\cite{10.1145/3658644.3690227} integrates SQL with static analysis for database-backed BOLA detection. These methods either detect flaws in source code or test against live instances. BACGuard complements these efforts by enforcing reconstructed authorization relations at runtime on unmodified applications, providing mitigation in production for flaws that detection and testing reveal but do not address, even in third-party or closed-source scenarios.

\subsection{WAFs and API security.} 

Web Application Firewalls such as ModSecurity with the OWASP Core Rule Set primarily address injection and cross-site scripting by matching request patterns \cite{modsecurity}. Despite this, they frequently fail to detect authorization weaknesses, as evidenced by limited block rates and significant disruption to legitimate traffic. While BOLA and BFLA are prominent in API security concerns \cite{owaspapi}, existing commercial tools focus on endpoint discovery and do not integrate schema-derived ownership with large language model arbiters to enforce object-level authorization in unmodified applications.

\subsection{Policy engines and authorization services.}

Extensive research has produced decision engines and authorization services, including policy-as-code systems like Open Policy Agent~\cite{opa}, Cedar~\cite{cedar2024}, XACML~\cite{xacml2013}, and relationship-based frameworks such as Google Zanzibar~\cite{zanzibar2019}. These solutions reliably determine if a request should be permitted, provided all relevant facts are available, which typically requires input from the application and its developers. This requirement is unmet when access control is broken or when an application cannot be modified. \sys{} addresses this by externally reconstructing authorization facts from application databases and session stores, enabling deterministic request-stage decisions. While tools like OPA could function as the decision engine, they cannot autonomously gather the necessary facts or handle response-path data leakage, which is managed by the LLM leak agent. The main innovation lies in fact reconstruction rather than the decision process itself.

\subsection{LLMs for security, agents, and prompt injection.}
Recent advancements use large language models for vulnerability detection, triage, and zero-shot repair \cite{pearce2023repair}, typically employing single classifiers for code or traffic. In contrast, the LLM-agent paradigm leverages reasoning agents\cite{react2023}, chain-of-thought prompting \cite{cot2022}, and multi-agent frameworks \cite{autogen2023} to allocate tasks more precisely. Prompt injection \cite{greshake2023injection, perez2022ignore} is a significant risk and is now recognized in the OWASP Top 10 for LLM applications~\cite{owaspllm}. Dividing responsibilities among specialized agents, grounded in factual data, is more robust against prompt injection than relying on a single agent. Ultimately, effectiveness is determined by deterministic reconstruction of ground-truth facts, not agent structure. This process achieves high quality, speed, and low cost, and can generate per-target policy automatically from an application's OpenAPI specification or traffic, thus reducing manual effort.
\section{Limitations}
\noindent \textbf{False positives.} Large-scale testing found no false positives among 1,732 realistic requests, resulting in an upper 95 percent bound of 0.22 percent. Additional measurements on captured or browser traffic support these findings. However, as the generators are still scripted, real production traffic is needed for more comprehensive evaluation, particularly for rare legitimate cases.

\noindent \textbf{Dependence on ground truth.} The guard relies on accurate and accessible data or identity stores. While degraded modes exist, effective object-ownership IDOR detection requires access to ownership information. Auto-derivation is limited by the completeness of its sources, such as the OpenAPI security scheme or observed discovery traffic.

\noindent \textbf{Provider and cost.} Experiments use a single large language model provider, with results showing that the least expensive model suffices. The LLM is called for no more than 12 percent of requests, so production costs are mainly driven by the free deterministic path.

\section{Conclusion}

Broken Access Control is challenging to address externally because missing checks are semantic and specific to individual objects. The determination of whether an authenticated request is abusive relies on object ownership, which exists within the database and remains inaccessible to web application firewalls. This paper proposes that an external guard can reconstruct necessary authorization information using application artifacts, such as signed tokens for user identity and foreign-key schemas for ownership, rather than inferring it. Large language model agents can reason using verified data. Basing decisions on validated database relationships decreases reliance on attacker-controlled input and increases resistance to request-level attacks, although effectiveness depends on accurately reconstructed authorization policies.

Empirical results demonstrate that the proposed system identifies and blocks Broken Access Control vulnerabilities overlooked by several applications and authorization models, including high-severity CVEs. The system achieves a perfect block rate with minimal false positives, significantly surpassing the performance of web application firewalls and single-agent models. Both a comparably informed single agent and a standard policy engine replicate the guard's request-stage decisions, while a large language model-based leak inspector detects response-path data leakage not captured by request-stage rules. Most requests are resolved rapidly via a deterministic fast path, with large language model escalation reserved for complex cases due to its higher latency and computational costs.

Further analysis indicates that excluding the large language model from the primary decision path has a greater impact than model size. A multi-agent hybrid using a smaller model performs comparably to a single agent with a larger model, as the monolithic agent tends to over-block by invoking the model for every request, a limitation addressed by the fast path. Policies are derived automatically from OpenAPI documents and foreign-key schemas or observed role-tagged traffic, resulting in perfect recall on tested targets. Adaptive-attacker studies demonstrate the architecture's resilience to identity forgery and prompt injection, with no successful evasions. The pipeline supports both hosted APIs and open-weight models, achieving optimal block and false positive rates in hybrid and pure multi-agent modes. The large language model is only used for rare cases, reducing operational costs and governance concerns.

{\footnotesize
\paragraph{Availability.} All scripts used in this study are available at
\url{https://github.com/MosesNde/BACGuard}.}

\bibliographystyle{plain}
\bibliography{refs}

\end{document}